# Reply to Commentaries on An Evolutionary Framework for Cultural Change: Selectionism versus Communal Exchange

Running head: Reply to Commentators: Evolutionary Framework Culture

LIANE GABORA
University of British Columbia

Pre-publication draft:
There may be minor differences between this draft and the final published version.

Address for Correspondence:
Liane Gabora
Department of Psychology, University of British Columbia
Okanagan Campus, Kelowna BC
Canada, V1V 1V7
E-mail: liane.gabora@ubc.ca
Phone: 604-822-2549 or 250-807-9849

The commentators have brought a wealth of new perspectives to the question of how culture evolves. Each of their diverse disciplines—ranging from psychology to biology to anthropology to economics to engineering—has a valuable contribution to make to our understanding of this complex, multifaceted topic. Though the vast majority of their comments were supportive of my approach, it is natural that a reply such as this focus on points where my views differ from that of the commentators.

The commentaries indicate strong support for the first basic premise of the paper: that ***culture does not evolve through a selectionist process***. There are two unsympathetic commentaries, one of which (**Simonton's**) offered no criticism, stating simply "space does not permit discussion here". The rest of this commentary, which summarizes his own theory of creativity, has little relevance to the content of the target paper.

The second unsympathetic commentary, **Madsen** and **Lipo's**, claims that my "primary mission is to "save culture" from being subject to natural selection". This is not the case; an examination of my papers on the topic over the last twenty years reveal the long and difficult struggle I went through trying to reconcile cultural change with the Darwinian perspective, initially believing that it was the only way *to* evolve.

**Madsen and Lipo** argue that culture evolves through a selectionist process on the basis of the following claims:

> "First, Gabora's argument is based upon a non-standard account of selection. Few if any geneticists or biologists would recognize genetic inheritance as powered by "self-replicating automata" of the von Neumann type. A "gene" (an incredibly complex concept in its own right [2]) certainly does not code both for the machinery to interpret itself, and for the protein its sequence represents."

I never claimed that a *gene* codes both for the machinery to interpret itself, and for the protein its sequence represents. I claimed that a *genome*—the *set* of all genes within an organism—constitutes a self-assembly code that is both *interpreted* during development and *copied* during reproduction, and this is standard biology; no living geneticist would argue with this.

**Madsen and Lipo** continue: "Second, Gabora adds unnecessary conditions to the requirements for selection." This is simply false, and the book by Holland that they cite at the end of this sentence says nothing about me, nor about unnecessary conditions to the requirements for selection. Instead of providing arguments to support their case they write:

> For example, the sequestration of inherited information, as in the isolation of germ-line cells in many animal lineages, is a derived characteristic which evolved quite late in the history of life [9]. Natural selection proceeds handily in many taxa which do not sequester heritable information.

*Natural selection* does not "proceed handily" in these taxa, but *evolution* does. Indeed that is exactly my point; it is *those* taxa evolve through, not natural selection, but communal exchange, and that served as the inspiration for the communal exchange theory of cultural evolution. I never said sequestration of inherited information is a requirement for *evolution*, I said it is a requirement for *evolution through natural selection*. This point is clearly understood by **Smith** (also an anthropologist), who writes, "As Carl Woese (1928-2012) pointed out, biology must abandon molecular reductionism and enter the nonlinear world [see reference 2]. Cultural evolution models must do the same, and Gabora's work is a significant step in the right direction." For decades it has been accepted that evolution through means other than natural selection is widespread. In addition to those cited in the target paper (such as [16, 25, 28]) thousands of other papers could be cited. As Koonin [17] puts it in a paper titled 'Darwinian evolution in the light of genomics':

> "Comparative genomics and systems biology offer unprecedented opportunities for testing central tenets of evolutionary biology formulated by Darwin in the *Origin of Species* in 1859 and expanded in the Modern Synthesis 100 years later. Evolutionary-genomic studies show that natural selection is only one of the forces that shape genome evolution and is not quantitatively dominant"

Dagan and Martin [4] suggest that as little as 1% of all species actually evolves as per a Darwinian 'tree of life' model; the rest exhibit the kind of network structure that characterizes cultural evolution as discussed in the target paper. Indeed in prokaryotes this kind of non-Darwinian horizontal exchange is so rampant that it appears that no two genes have the exact same evolutionary history [2, 3, 13, 18, 19, 30, 31].

**Madsen and Lipo** also claim, "Nor is it required that new variation be approximated by a random process." If the explanation provided in the target paper was not sufficiently clear, I encourage **Madsen and Lipo** to consult any basic population genetic text (such as Hartl and Clark's [14]) to understand why natural selection entails random variation. Not only does generation operate *prior to* selection, but it can change the distribution of variants instantaneously, whereas selection requires generations to do so. The speed and effectiveness of a generation bias renders any selection bias negligible. Generation biases exist in the biological world (e.g., mutation biases or nonrandom mating), but to a degree that is miniscule compared with culture, where novelty is not just "biased by" but strategically designed to meet human needs, tastes, and preferences.

The logic of this is seen clearly when one writes computer programs to simulate natural selection (as I once did as the teaching assistant for a population genetics class). Natural selection operates when some traits make it harder for organisms that bear them to survive and reproduce, so that there are fewer and fewer organisms with those traits in subsequent generations. The greater the extent to which variants are *generated* non-randomly, the greater the extent to which what you end up with over time reflects *not* selection (which occurs *after* variants have come into existence) but the *initial generation bias*. To take a cultural example, the greater the extent to which, say, a novel is generated non-randomly, the greater the extent to which the wording and organization of novels reflect the thought processes of their authors. **Madsen and Lipo** do not provide any reason for their conviction that variation need not be random, nor do they address the arguments on this matter in the target paper. They merely cite a paper by anthropologists [15] that describes a model in which variation is generated through random copying error, i.e., a paper that in no way addresses their claim that natural selection can be operative when variation is non-random.

**Madsen and Lipo** continue:

> "Third, Gabora criticizes the transfer of biological concepts to culture, by confusing theoretical concepts and empirical instances. For example, she claims that

> "generations" are not applicable to culture. In reality, organismal generations are an empirical matter, and depend upon the details of a transmission process. For culture, the lack of a convenient empirical package to identify generations simply points to the need to define and model appropriate time scales for analysis.

I am all for defining and modeling appropriate time scales for analysis, but that is irrelevant; in a realm where entities can appear to be 'dead' and then suddenly resume the characteristics of life, and do so regularly, the concept of 'generation' *does* break down. **Madsen and Lipo** then say: "Such a challenge is also routine in biology when germ-line sequestration is not present or reproduction is age-structured or reproduction continuous". I agree; again, it was through my study of these problems in the microbial world that it became evident that they were problematic with respect to cultural evolution.

Finally, **Madsen and Lipo** write:

> Using a well-accepted definition of natural selection [12], it is apparent that any population in which individuals vary in their phenotype, **inherit** the information underlying that phenotype, and where the success of individuals is affected by that **heritable** information, will evolve. Observed changes in the frequencies of heritable cultural information constitute natural selection, given these conditions. [bold added]

It is ironic that they cite Lewontin here given his well-known critique of Darwinian approaches to culture [5]. In the situation they describe, information is not acquired but inherited, and therefore natural selection *may* be taking place. Nothing I have ever said contradicts this. However, they continue:

> In this light, Gabora's EVOC model [1] displays all the hallmarks of natural selection, demonstrating that "communal exchange" is a viable model of an inheritance and selection process, not a replacement for selection.

EVOC models an evolutionary process, but it exhibits characteristics that disqualify it as evolution through natural selection, such as, (1) information is acquired not inherited, and (2) variants are generated non-randomly, as discussed above, and discussed more extensively in the target paper. Thus it does not support the idea that communal exchange is an "inheritance and selection process". As Lynch [20] puts it:

> It has long been known that natural selection is just one of several mechanisms of evolutionary change, but the myth that all of evolution can be explained by adaptation continues to be perpetuated by our continued homage to Darwin's treatise (6) in the popular literature. For example, Dawkins' (7–9) agenda to spread the word on the awesome power of natural selection has been quite successful, but it has come at the expense of reference to any other mechanisms, a view that is in some ways profoundly misleading.

I agree with **Smith** that culture "retains a place for selection" in the sense that individuals select one brand of peanut butter over another. This is quite different from selection in the sense of a selectionist algorithm, i.e., selection of certain instruction sets that code for self-replicating structures at the expense of others. I agree with **Orsucci** that an important related project is to study the co-evolution of cultural and genetic systems, and many including myself have taken steps in this direction. This has always been an important component of the communal exchange approach to cultural evolution, and although it is not developed extensively in this paper, it is an area ripe for future work. Establishing an adequate theoretical framework for cultural evolution will provide a natural foundation for work in this area. If in the future cultural evolution experiences (as did biological evolution did several billion years ago) what Vetsigian et al. [28] refer to as a Darwinian threshold, a transition from a communal exchange to a selectionist evolutionary process then, as **Orsucci** suggests, communal exchange and selectionist theories would be complementary components of a cohesive theory of cultural evolution.

**Orsucci** also points out that, “A focus on artifacts and artistic expressions could lead to difficulties related to their iconic nature, in semiotic terms, and the identification of markers for underlying codes might be left to inferential and subjective approaches.” Fortunately, the need for inference is decreasing as neuroscience progresses toward the point where thoughts about artifacts and artistic expressions can be reliably connected to underlying patterns of neural activation.

**Communal Exchange Theory**

The commentaries also indicate strong support for the second basic premise of the paper: ***culture evolves through communal exchange amongst self-organizing, autopoietic networks, in a manner that is algorithmically similar to metabolism-first theories of the origin of life***. **Smith** points out that the communal exchange approach to culture “avoids a monolithic conception of ‘culture’ (again a common item of critique in evolutionary approaches to culture)” and that it “places a premium on understanding the flow of cultural information communicates well with such traditional cultural-anthropological subjects as power relationships”. **Kauffman** points out that the fact that elements of culture do not evolve as separate entities but as facets of an integrated whole (a worldview) is particularly in evidence with respect to science, in that a scientific hypothesis builds upon (and may fall with) other scientific hypotheses. **Kauffman** also notes that enabling the user of the conceptual network program for documenting material cultural history to “introduce context-relevant attributes and perspectives is of particular importance, and should be highlighted,” because this shows “implicit acceptance that the phase space is not pre-state-able”. Indeed, the conceptual network program is part of a family of approaches referred to collectively as “human computation” systems [21] that tackle the kind of intractability Kaufman describes by merging the inductive abilities of individuals and crowds with the speed and accuracy of computers at algorithmically solvable tasks.

One commentator (**Brown**) is in agreement with my critique of Darwinian theories of culture but unsympathetic with the view that culture evolves by *some* means. He claims,

> “The search for the “right” evolutionary framework for cultural change is obviously motivated by the belief that culture is something that “evolves.” … But how may cultural change be described as adaptive? … [I]n what sense might the proliferating use of video games or tanning salons be classified as adaptive cultural change?”

It is true that video games and tanning salons may not be adaptive for humans evolving on a *biological* fitness landscape, but that is not relevant to their cultural evolution, and indeed the two are often at odds. A cultural adaptive landscape need not be defined by natural laws; all that is necessary is that there exist some criterion according to which some outputs rate higher than others. This rating can be in terms of number of offspring, as in biology, or it can be defined by human-made rules, preferences, or even dollars. As **Smith** notes, the communal exchange approach, "highlights the importance of individual agency in reshaping culture moment by moment".

**Brown** also claims,

> "An adaptive genotypic trait gives survival advantages to *all* members of the species that possess it. By contrast, a successful—meaning broadly rooted—cultural innovation (such as intellectual property rights) may advantage some individuals or groups and disadvantage others."

**Brown** is mistaken that biological and cultural evolution differ in this respect. There are genetic traits that benefit some individuals and that are maladaptive for, detrimental to, or even not expressed at all in, others. A classic example of this is the gene that causes sickle-cell anaemia in homozygous individuals (those who bear two copies of the gene), but confers resistance to malaria in heterozygous individuals (those who bear one copy of the gene). As another example, genes that make an individual exhibit a sudden growth spurt during adolescence are beneficial for basketball players but not for gymnasts.

**Bejan** suggests that development of a communal exchange framework for cultural evolution falls under the umbrella of *construal law*, a program that investigates the physical rules underlying designs, both natural and human-made. Some might fear that use of the word "design" implies a "designer" but as I understand it, Bejan uses the term to refer to form that emerges or evolves through low-level physical processes as well as form that comes about through the actions of intentional agents. I am sympathetic toward attempts to develop this kind of unified framework, and my colleagues and I have made similar efforts [1, 9, 10, 11]. Bejan's framework is interesting, though in my view somewhat too simplified with respect to its characterization of living, cognitive system to capture their essential features. For example, there is much more to being alive than "to persist in time" (indeed a dead body continues to "persist in time", and non-living systems often persist substantially longer than living ones). As another example, **Bejan's** explanation for how physics can explain even the existence of the peacock's tail invokes "mating", which is not the domain of physics. Nevertheless his commentary demonstrates that by viewing biological and cultural systems through the lens of construal theory one gains a refreshingly new and valuable perspective.

**Simonton** outlines a theory of creativity that ignores "the minor complications introduced when selection is sequential rather than simultaneous". However, selection *by definition* entails that entities are simultaneously chosen amongst; selectionist processes are subject to order effects such that if some are selected before others, they change the fitness landscape by which the others are evaluated. Thus a selectionist model by definition cannot describe sequential change. Moreover, sequential change is not a "minor complication"; it necessitates a different mathematical structure (for discussion, see [6, 7, 8]). Furthermore, sequential refinement of ideas over time is the rule, not the exception. Indeed the exception to this rule, the only sort of creativity that a model such as Simonton's could hope to describe, is insight, where an idea manifests suddenly and unexpectedly. However insight would seem to be more aptly described as reflecting a phase transition, i.e., a sudden re-organization of knowledge due to self-organized criticality [6, 24], rather than an executive level selection amongst competing alternatives.

Another problem with Simonton's theory of creativity is that it starts from the assumption that "the individual produces a set $X$ containing the potential solutions $x_1$ ... $x_i$ ... $x_k$, where $k \geq 1$". It is not explained how different $x_1$ has to be from $x_i$ to be considered a different solution; indeed it is generally the case that different possibilities constitute a family of interrelated solutions. Without going into the details, this interrelatedness is why in the honing theory of creativity they are characterised as different *actualizations* of a core idea, which exists, initially, in a "ground state", or state of potentiality [6, 7, 12].

I conclude by saying that I am grateful to the commentators for their diverse perspectives and insights, their overall support for the project, and provocative ideas for where to go from here. Clearly there are many fascinating avenues to explore as we move forward on our quest to understand how culture evolves.

**Acknowledgements**
This research was conducted with the assistance of grants from the National Science and Engineering Research Council of Canada, and the Fund for Scientific Research of Flanders, Belgium.

**References**

1. Aerts, D., Bundervoet, S., Czachor, M., D'Hooghe, B., Gabora L., & Polk, P. On the foundations of the theory of evolution. In M. Locker (Ed.) *Systems Theory in Philosophy and Religion, Vols I & II*. Windsor, Canada: IIAS, 2006.
2. Andam C. P., Gogarten J. P. Biased gene transfer in microbial evolution. Nature Reviews Microbiology 9 (2011) 543-555.
3. Boto L. Horizontal gene transfer in evolution: facts and challenges. Proceedings of the Royal Society: Biological Sciences 277 (2010) 819-827.
4. Dagan, T., Martin, W. *Genome Biology* **7** (2006) e118.
5. Fracchia J., & Lewontin, R. C. Does culture evolve? *History and Theory, 38* (1999) 52–78.
6. Gabora, L., Cognitive mechanisms underlying the origin and evolution of culture. Doctoral thesis, Center Leo Apostel For Interdisciplinary Studies, 2001.
7. Gabora, L., Creative thought as a non-Darwinian evolutionary process. Journal of Creative Behavior, 39 (2005) 65–87.
8. Gabora, L. An analysis of the Blind Variation and Selective Retention (BVSR) theory of creativity. *Creativity Research Journal, 23* (2011) 155-165.
9. Gabora, L., & Aerts, D. Distilling the essence of an evolutionary process, and implications for a formal description of culture. In (W. Kistler, Ed.) *Proceedings of the Center for Human Evolution: Workshop 4 on Cultural Evolution*, Foundation for the Future, Belmont WA, 2005.
10. Gabora, L., & Aerts, D. Evolution as context-driven actualization of potential: Toward an interdisciplinary theory of change of state. *Interdisciplinary Science Reviews, 30* (2005) 69-88.
11. Gabora, L., & Aerts, D. A cross-disciplinary framework for the description of contextually mediated change. In (I. Licata & A. Sakaji, Eds.) *Physics of Emergence and Organization,* (pp. 109-134). Singapore: World Scientific, 2008.

12. Gabora, L., & Saab, A. Creative interference and states of potentiality in analogy problem solving. Proceedings of the Annual Meeting of the Cognitive Science Society (pp. 3506-3511). July 20-23, Boston MA, 2011.
13. Gogarten J. P., Townsend J. P. Horizontal gene transfer, genome innovation and evolution. Nature Reviews Microbiology 3 (2005) 679-687.
14. Hartl, D. L. & Clark, A. G. Principles of Population Genetics (Sinauer Associates, Sunderland, MA, 1997).
15. Henrich, J., Boyd, R. On modeling cognition and culture: Why cultural evolution does not require replication of representations. Journal of Cognition and Culture 2 (2002) 87–112.
16. Koonin E. V., Makarova K. S., Aravind L. Horizontal gene transfer in prokaryotes: quantification and classification. Annual Review of Microbiology 55 (2001) 709-742.
17. Kauffman, S., Origins of order. New York: Oxford University Press, 1993.
18. Koonin, E. V. Darwinian evolution in the light of genomics. Nucleic Acids Research, 37 (2009) 1011–1034.
19. Koonin, E. V. and Wolf, Y. I. Evolution of microbes and viruses: A paradigm shift in evolutionary biology? Frontiers in Cellular and Infection Microbiology, 2 (2012) 119.
20. Lynch, M. The frailty of adaptive hypotheses for the origins of organismal complexity. Proceedings of the National Academy of Science USA, 104 (2007): 8597–8604.
21. Michelucci, P. (Ed.) *Handbook of Human Computation*. (in press.) Springer.
22. Pigliucci, M. An extended synthesis for evolutionary biology. Annals of the New York Academy of Sciences, 1168 (2009) 218-28.
23. Raoult, D. The post-Darwinist rhizome of life. Lancet 375 (2010) 104–105.
24. Ward, T. B., Smith, S. M., & Vaid, J. (1997). Conceptual structures and processes in creative thought. In Ward, T. B., Smith, S. M. & Vaid. J. (Eds.) Creative Thought: An investigation of conceptual structures and processes (pp. 1-27). Washington, DC: American Psychological Association.
25. Woese, C. R., On the evolution of cells. Proceedings National Academy Science, 99 (2002) 8742–8747.
26. Woese C. R., Goldenfeld N. How the microbial world saved evolution from the scylla of molecular biology and the charybdis of the modern synthesis. Microbiology and Molecular Biology Reviews 73 (2009) 14–21.
27. Woese, C. R., Olsen, G. J., Ibba, M., and Soll, D. Aminoacyl-tRNA synthetases, the genetic code, and the evolutionary process. Microbiology and Molecular Biology Reviews 64 (2000) 202–236.
28. Vetsigian, K., Woese, C., & Goldenfeld, N., Collective evolution and the genetic code. Proceedings National Academy Science, 103 (2006) 10696–10701.
29. Villarreal, L. P., and Witzany, G. Viruses are essential agents within the roots and stem of the tree of life. Journal of Theoretical Biology 262 (2010) 698–710.
30. Zhaxybayeva O. Detection and quantitative assessment of horizontal gene transfer. Methods in Molecular Biology 532 (2009) 195–213.
31. Zhaxybayeva O., Doolittle W. F. Lateral gene transfer. Current Biology 21 (2011) R242-R246.